\documentclass{aa}
\usepackage{graphicx}
\usepackage{txfonts}
\newcommand{\NaI}{\ion{Na}{I}}
\newcommand{\KI}{\ion{K}{I}}
\newcommand{\CrI}{\ion{Cr}{I}}
\newcommand{\CaI}{\ion{Ca}{I}}
\newcommand{\MnI}{\ion{Mn}{I}}

\begin{document}

\title{VLT/UVES spectroscopy of 
    V4332 Sagittarii in 2005: \\ 
    The best view on a decade-old stellar-merger remnant}

\author{R. Tylenda\inst{1}, S. K. G\'orny\inst{1}, 
T. Kami\'{n}ski\inst{2}, and M. Schmidt\inst{1}}

\institute{Department of Astrophysics, 
       Nicolaus Copernicus Astronomical Center,                   
       Rabia\'{n}ska 8,
       87-100 Toru\'{n}, Poland\\
       \email{tylenda@ncac.torun.pl}
    \and
       European Southern Observatory, Alonso de C\'ordova 3107, Vitacura,
       Santiago, Chile}

\date{Received; accepted}

\abstract
{The source V4332 Sgr is a red transient (red nova) whose eruption was observed in
1994. The remnant of the eruption shows a unique optical spectrum: strong
emission lines of atoms and molecules superimposed on an M-type stellar
spectrum. The stellar-like remnant is not directly observable,
however.
It is presumably embedded in a disc-like dusty envelope seen almost
face-on. The observed optical spectrum is assumed to result from 
interactions of the central-star
radiation with dust and gas in the disc and outflows initiated in 1994.}
{We aim at studying the optical spectrum of the object in great detail
to better understand the origin of the spectrum and the nature of the object.}
{We reduced and measured a high-resolution (R $\simeq 40\,000$) spectrum 
of V4332~Sgr obtained with VLT/UVES in April/May 2005. 
The spectrum comes from the ESO archives and
is the best quality spectrum of the object ever obtained.}
{We identified and measured over 200 emission features belonging to
11 elements and 6 molecules. The continuous, stellar-like component
can be classified as $\sim$M3. The radial velocity of the object is $\sim -75$ ~km\,s$^{-1}$
as
derived from narrow atomic emission lines. The
interstellar reddening was estimated to be $0.35 \le E_{B-V} \le 0.75$.
From radial velocities of interstellar absorption features in the \NaI\,D lines,
we estimated a lower limit of $\sim$5.5~kpc to the distance of V4332~Sgr.
When compared to
spectroscopic observations obtained in 2009, the spectrum of V4332~Sgr considerably 
evolved between 2005 and 2009. The object significantly faded in the optical 
(by $\sim$2 mag in the $V$ band), which resulted from cooling
of the main
remnant by 300--350~K, corresponding to its spectral-type change 
from M3 to M5-6. 
The object increased in luminosity by $\sim 50$\%, however,
implying a
significant expansion of its dimensions.
Most of the emission features seen in 2005 significantly faded or even disappeared
from the spectrum of V4332~Sgr in 2009. These resulted from
fading of the optical central-star radiation and a decrease of the
optical thickness of the cirumstellar matter, presumably due to its
expansion.
V4332 Sgr bears several resemblances to V1309~Sco, which erupted in 2008.
This can 
indicate a similar nature of the eruptions of the two objects.
The outburst resulted from merger of a contact binary in V1309 Sco.}
{}


\titlerunning{Optical spectrum of V4332 Sgr}
\authorrunning{Tylenda et al.}
\maketitle
       

\section{Introduction \label{intro}}

The source V4332 Sagittarii (V4332 Sgr) was discovered as Nova Sgr 1994 in
February~1994 \citep{hayashi}. 
Spectroscopic observations showed narrow Balmer lines in
emision superimposed on a K-type (super)giant spectrum, however,
that quickly evolved to
M type (super)giant \citep{tomaney,martini}. 
This spectral apearance and evolution was at variance with what is observed 
in classical novae, but bore a resemblance to the luminous
red variable observed in M31 in 1988 (M31\,RV) \citep{mould}. At present,
stellar eruptions of this type are called red transients,
 intermediate-luminosity red transients, red novae, 
 luminous red novae, or V838 Mon-type objects. 
The latter name comes from the gigantic eruption of
V838~Monocerotis observed in 2002 \citep{muna02,crause03}, which elicited
great interest in astrophysicists, as well as public media, partly due to 
the spectacular light-echo event accompanying the outburst \citep{bond03}.

In addition to these three objects, the class of red transients in
our Galaxy
includes V1309~Scorpii (V1309~Sco) \citep{mason10} and OGLE-2002-BLG-360
\citep{tku13}. V1148~Sagittarii (Nova~Sgr~1943)
probably also belonged to this class, as can be inferred from its spectral
evolution described by \cite{mayall}. There is also growing observational 
evidence that CK~Vulpeculae (CK~Vul, Nova~Vul~1670) \citep{shara85} was 
a red transient and not a classical nova \citep{kato,tku13,kmt15}.
A few extragalactic objects, usually referred to as intermediate-luminosity
optical transients, for instance, M85\,OT2006 \citep{kulk07}, NGC300\,OT2008
\citep{bond09,berger09}, and SN~2008S
\citep{smith09}, might be of a similar nature.

Although they are different in light curve, time scale, and peak
luminosity, red transients always show a similar spectral evolution: in
course of the eruption, the objects evolve to progressively lower effective
temperatures and decline as M-type (super)giants. Their remnants also
resemble a late M-type (super)giants with a significant (often dominating)
infrared excess.

Several mechanisms have been proposed to explain the red-transient events, 
including an unusual nova 
\citep{it92}, a late He-shell flash \citep{law05}, and a stellar
merger \citep{soktyl03}.
They have been critically discussed in \cite{tylsok06}.
These authors concluded that the only mechanism that
can satisfactorily account for the observational data of red transients is 
a merger of two stars. For the case of V838~Mon they argued that this
eruption might have been due to a merger of a low-mass pre-main-sequence star 
with an $\sim 8\,M_\odot$ main-sequence star.

The source V1309 Sco, which erupted in 2008 \citep{mason10}, appeared to be a sort of
Rosetta stone for understanding the nature of red transients. Thanks to the
archive data from the Optical Gravitational Lensing Experiment (OGLE)
\citep{udal03}, it was
possible to follow the photometric evolution of the object during six years
before the outburst \citep{thk11}. The result was amazing: the progenitor
of V1309~Sco was a contact binary that quickly lost its orbital angular
momentum and evolved into a merger of the components. Thus V1309~Sco
provided strong evidence that the red transients are indeed caused
by stellar mergers.

After the 1994 outburst, V4332 Sgr was almost forgotten by astrophysicists and
observers. No observations of the object were made until 2002, 
when V838~Mon erupted and astronomers realized that
these two objects most probably belong to the same class. Then V4332 Sgr regained
astrophysical interest. Several spectroscopic observations made in
2002-2003 revealed an unusual
optical spectrum for stellar objects: strong and numerous emission lines 
of atoms and molecules were superimposed on a weak, early-M-type stellar
spectrum \citep{baner,tcgs05,kimes}. In addition,
strong bands of AlO in emission were detected in the near-IR
spectral region \citep{baner03}.

A detailed study of the emission-line spectrum and the spectral energy
distribution (SED) of V4332~Sgr led \cite{kst10} to conclude that the main
remnant of the 1994 eruption is now obscured to us, most probably the
central object is embedded in a dusty disc seen almost edge-on.
The observed optical spectrum is assumed to be produced by interactions 
of the central star's radiation with the matter in the disc and the outflows
originating from the 1994 eruption: 
the M-type continuum results from scattering on dust grains, 
while the emission-line spectrum is due to resonant scattering 
by atoms and molecules. This conclusion was subsequently confirmed by
polarimetric \citep{kt11} and spectropolarimetric \citep{kt13} observations,
which showed that the optical continuum is strongly polarized, while the
emission features are mostly unpolarized.

\begin{figure}
  \includegraphics[width=\columnwidth]{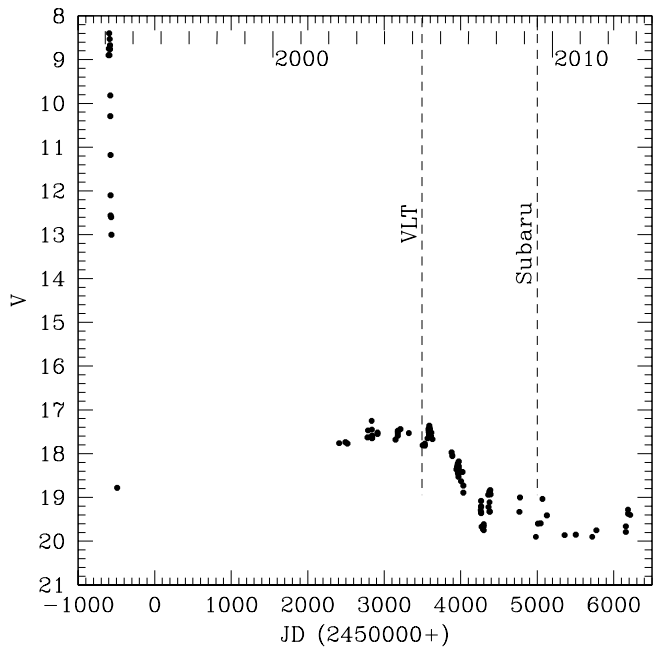}
  \caption{Light curve of V4332~Sgr in the $V$ band. Vertical dashed
lines indicate the time moments of the VLT spectroscopy we analysed and the observations obtained with the Subaru telescope in June 2009 
that are described in \citet{kst10}.}
\label{lc_fig}
\end{figure}

We present an optical spectrum of V4332~Sgr obtained
in 2005. The data come from the archives of the Very Large
Telescope (VLT) and were not published before. 
The quality and resolution of the
spectrum is exceptional; the data present the best quality spectrum of
V4332~Sgr ever obtained. Since 2005, the object has significantly faded 
(see Fig.~\ref{lc_fig}) and its spectrum
has considerably evolved \citep[see e.g.][]{barsuk}. 
Therefore the spectrum described and analysed in 
the next sections presents a unique set of data
on V4332~Sgr, which
was the main reason for us to reduce and publish the data. 

\begin{figure*}
   \includegraphics[width=\textwidth]{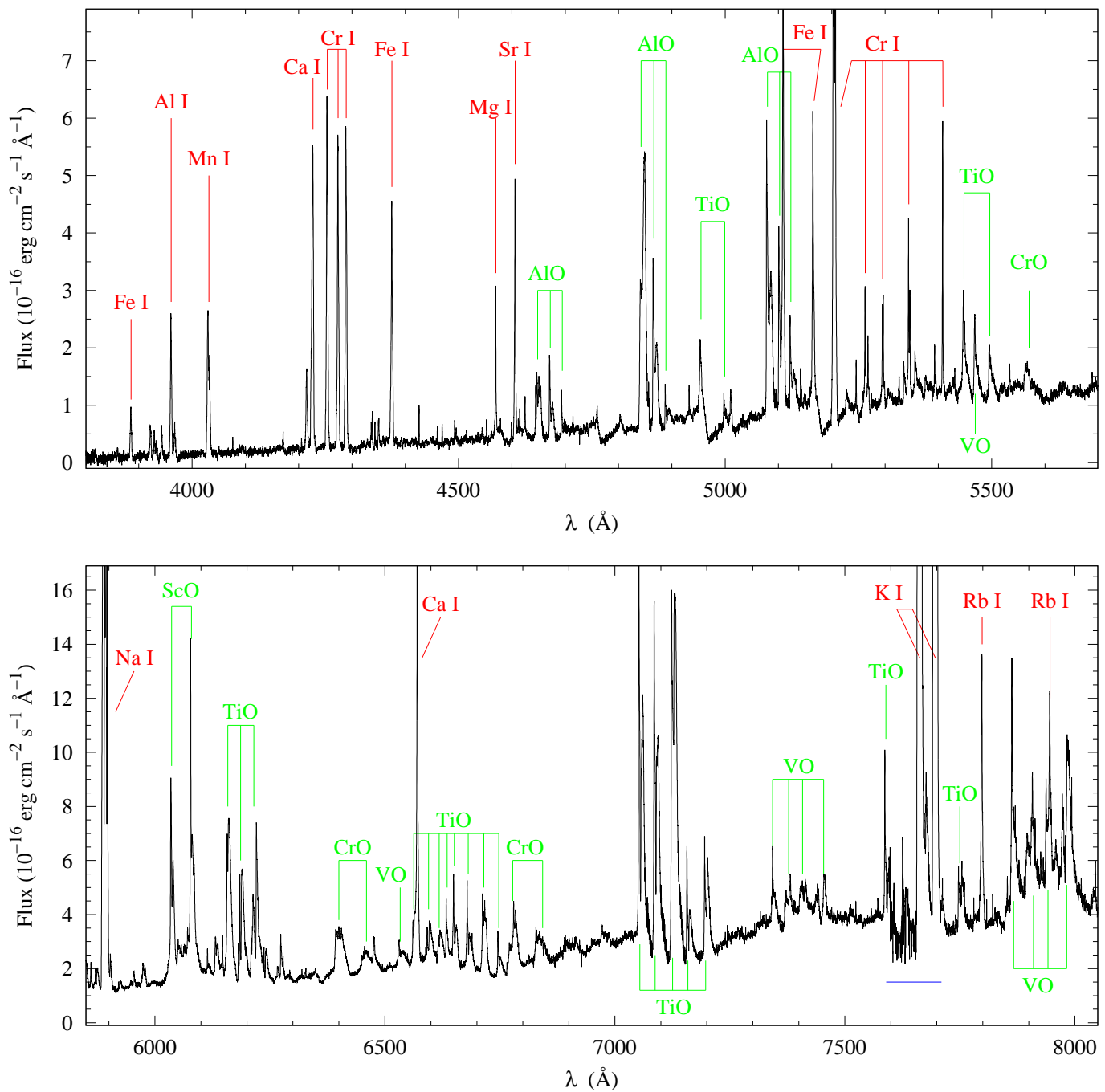}
   \caption{Spectrum of V4332~Sgr obtained in 
     April--May~2005 with the UVES/VLT. 
     The displayed spectrum was smoothed from the original resolution with
     boxcar~10. The strongest atomic (red) and molecular (green)
spectral
     features are indicated. 
     A blue horizontal bar indicates the spectral
     region affected by telluric absorption bands.}
     \label{spec}
\end{figure*}

\section{Observations and data reduction \label{obsred}}

We have found high resolution spectra of 
V4332~Sgr in the ESO data archives. They have been carried out with UVES/VLT in 2005
on 22~April and 12~May. The observations were made in the framework of 
the 075.D-0511 (PI: Banerjee) observing programme.
The spectra were obtained with three different 
spectrograph settings covering
the range 3756--10253~\AA\ with two gaps at 5750--5833~\AA\ and
8520--8656~\AA. The technical details of the observations 
are provided in Table~\ref{log_tab}.

Figure \ref{lc_fig} shows the light curve of V4332~Sgr in the $V$
photometric band since the discovery of the object in February~1994. The
data are taken from the compilation of
V.~Goranskij\footnote{\tt http://jet.sao.ru/$\sim$goray/v4332sgr.ne3}
\citep[see][for the light curves in different photometric bands]{barsuk}. 
Two vertical dashed lines
indicate the time moments of the VLT spectroscopic observations we analysed and the observations obtained with the Subaru telescope
in June~2009 that were presented in \citet{kst10}.

One blue and two different red standard UVES configurations were
used to observe the target.  The observations centred on 8600\AA\ initially
split into two exposures were later merged.  Since in the case of the red
configurations there are two separate lower and upper parts, the final
spectra consist of five independently registered elements.  The CCDs were
read with 2x2 binning in all configurations.  
The total useful spectral range covers 3760-9500\AA,\ with
two small overlapping regions of individual sections and some gaps between
lower and upper parts of the red configurations.

\begin{table*}
\begin{center}
\caption{ Log of observations of V4332 Sgr with UVES/VLT  \label{log_tab}}
\begin{tabular}{ c l c c c }
\hline
 date \& time        & configuration  & range(\AA)   & resolution & exp. time
\\
\hline
 2005.04.22 06:22:01 & RED 580        & 4727~~--~~6835 & 42310 & 3000 sec   \\
 2005.05.12 06:10:19 & RED 860        & 6650 -- 10250 & 42310 & 1480 sec   \\
 2005.05.12 06:10:23 & BLUE 437       & 3730~~--~~5000 & 40970 & 2960 sec   \\
 2005.05.12 06:43:16 & RED 860        & 6650 -- 10250 & 42310 & 1100 sec   \\
\hline
\end{tabular}
\end{center} 
\end{table*}

From the ESO data archive we have downloaded the spectra in their raw, 
unreduced form together with
the full set of reduction and calibration files for each of the standard
configurations.
The reduction and calibration of the spectra was made with
the ESO-MIDAS reduction software.

The background was subtracted from flat-field, arc, and
science frames. 
The Th-Ar lamps were used for wavelength 
calibration.
Flat-fielding was performed in the pixel-pixel space.  
The signs of cosmic rays and other defects of the CCD were removed from each  
science frame using the standard MIDAS procedures.
The extracted
spectra were wavelength calibrated and corrected for atmospheric extinction.
The flux calibration was performed using the master response calibration    
files prepared by ESO for each of the standard UVES configurations.  
Telluric lines were partially removed using {\tt Molecfit} \citep{molecfit}.
 The procedure left residuals below 20\% of the original strength of the telluric lines.
The individual heliocentric velocity corrections were
applied to each individual spectrum of V4332~Sgr.  
As a last step, the five parts were combined into a single
file, averaging the overlapping regions.

\section{Spectrum \label{results}}

The final flux-calibrated 1D spectrum of V4332 Sgr is  
presented in Fig.~\ref{spec}. The spectrum above $\sim$8100\,\AA\ is not
shown in the figure because it is heavily
disturbed by telluric absorption lines and only shows a few emission
features of the object (see Table~\ref{molec_tab}).
The wavelength is given  in the heliocentric
rest frame, and the flux is in units of 
$10^{-16}$~erg~s$^{-1}$~cm$^{-2}$~\AA$^{-1}$.
The strongest atomic and molecular features in emission are
indicated in the figure.

The atomic line identification was based mainly on the NIST Atomic Spectra
Database \citep{nist}\footnote{\tt 
http://physics.nist.gov/asd}, the Atomic Line List v2.04 by
van Hoof\footnote{\tt http://www.pa.uky.edu/\-$\sim$peter/\-atomic/},
and on the multiplet tables by \cite{moore}. 
All the identified atomic lines seen in emission in the spectrum of V4332~Sgr 
are listed in Table~\ref{atom_tab}. The observed wavelengths (in \AA) 
of the lines
are given in Col.~(1) of the table, while their laboratory wavelengths and
ion identification can be found in Cols.~(2) and (3), respectively.
Column~(4) presents the measured fluxes (in units of
$10^{-16}$~erg~s$^{-1}$~cm$^{-2}$) of the lines with their estimated
uncertainties. Column~(5) gives the full widths at half maximum (FWHM in
\AA) of the lines. The last column lists notes on
the lines and their measurements. The explanations of the symbols used 
in Table~\ref{atom_tab} are given in Table~\ref{notes_tab}.

All the identified molecular bands seen in emission in the spectrum of 
V4332~Sgr are listed in Table~\ref{molec_tab}. The method used to identify the
bands with appropriate references for molecular data can be found in 
\citet{kst09} and \citet{kst10}. The first column of the table gives
wavelengths of the bands. These are mostly values from
laboratory measurements and refer to the heads of the bands. In a few cases,
the wavelengths are results of theoretical estimates. The identification of the
bands is provided in Col. (2), while Col. (3) displays the measured
fluxes (in $10^{-16}$~erg~s$^{-1}$~cm$^{-2}$) in the bands
with their accuracies. The last column gives notes on 
the bands and their measurements. The symbols used
in these two columns are explained in Table~\ref{notes_tab}.

\section{Spectral classification of the stellar-like continuum \label{sp_class}}

To estimate the spectral type of the stellar-like
continuum observed in V4332~Sgr, we attempted to fit standard and model 
spectra to the results of our observations.
Standard stellar spectra were taken from \citet{jacob}, while model atmosphere
spectra were obtained using the MARCS grid \citep{marcs}.

\begin{figure}
  \includegraphics[width=\columnwidth]{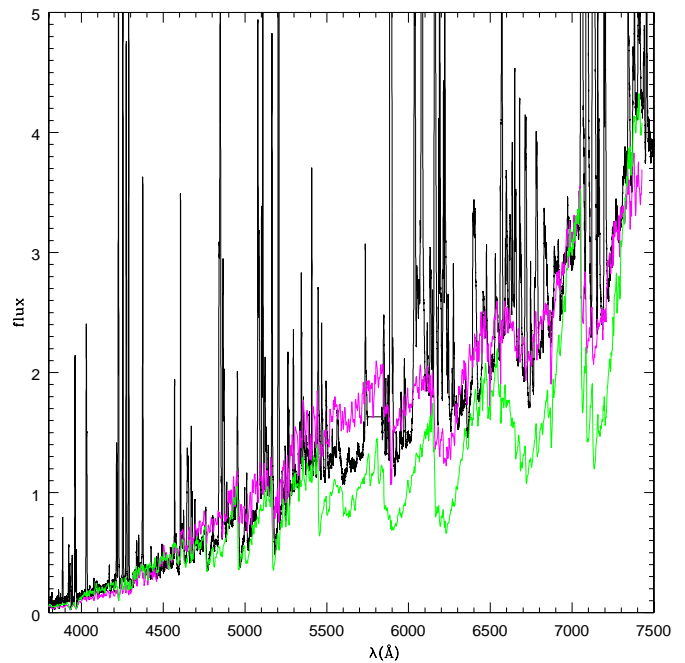}
  \caption{Standard giant spectra fitted to the observed spectrum of
  V4332~Sgr (black). Magenta: M2 standard spectrum reddened with 
   $E_{B-V} = 0.4$. Green: M4 standard spectrum reddened with 
   $E_{B-V} = 0.15$.}
\label{sp_class_1_fig}
\end{figure}

Figure~\ref{sp_class_1_fig} shows M2 and M4 giant spectra fitted to the
observations. An interstellar reddening was applied to the standard spectra
to obtain a good fit at the short- and long-wavelength edges of 
the spectra. These were $E_{B-V} = 0.4$ for the M2 and 0.15 for the M4 spectrum. As can be seen from
Fig.~\ref{sp_class_1_fig}, neither of the two standard spectra satisfatorily
fits the observations. The M2 spectrum presents absorption bands somewhat too 
shallow when compared to the observed continuum, while
its flux in the middle spectral region, that is, between $\sim$5000\,\AA\ and 
$\sim$6500\,\AA, is systematically too high. In contrast, the absorption bands of the M4 spectrum are too deep, and its flux in the middle region is too low.
The observed continuum generally lies in between the M2 and M4 standard
spectra, suggesting that an M3 giant spectral type probably
reproduce the observations adequately.

\begin{figure}
  \includegraphics[width=\columnwidth]{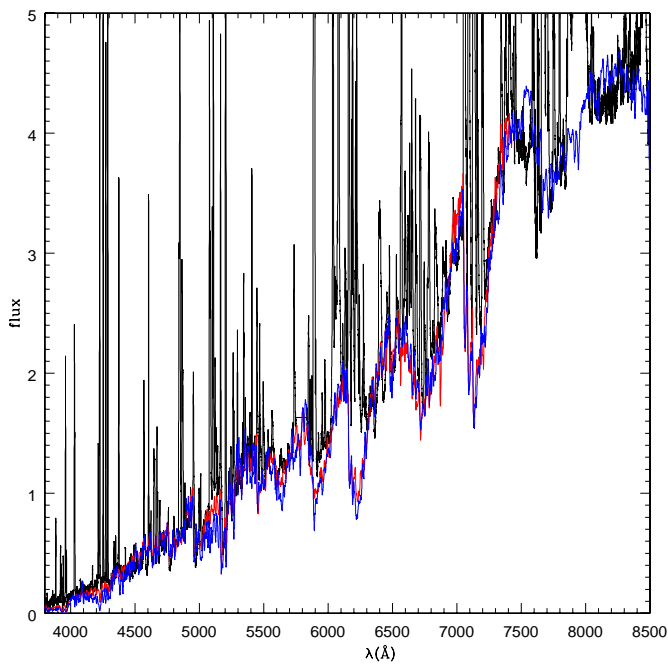}
  \caption{Standard and model giant spectra fitted to the observed spectrum of
  V4332~Sgr (black). Red: M3 standard spectrum reddened with 
   $E_{B-V} = 0.35$. Blue: model atmosphere ($T_{\rm eff} = 3600$~K, 
   log~$g$ = 0.5) spectrum reddened with $E_{B-V} = 0.35$.}
\label{sp_class_2_fig}
\end{figure}

Indeed, as shown in Fig.~\ref{sp_class_2_fig}, the M3 standard spectrum,
as well as the MARCS model spectrum calculated for stellar parameters typical
for an M3 (super)giant
\citep[$T_{\rm eff} = 3600$~K, log~$g = 0.5$,][]{levesque},
fit the observed continuum. Some absorption bands, for example that at
$\sim$6200\,\AA,\, may seem to be too deep in the reference spectra compared to the observed band. However, in the spectrum of V4332~Sgr these
absorption features are partly filled by strong, partly overlapping emissions. 
Note that the two reference spectra
shown in Fig.~\ref{sp_class_2_fig} were
reddened with $E_{B-V} = 0.35$.

\section{Interstellar reddening  \label{reddening}}
 
The interstellar extinction towards V4332~Sgr was estimated
in a number of papers \citep{martini,tcgs05,kimes}. 
These results as well as their own determinations were discussed in
\citet{kst10}. Most of the estimates relied on a comparison of the observed
continuum of V4332 Sgr with stellar standards. \citet{kst10} finally adopted
$E_{B-V} = 0.32$, although the values varied between 
0.22 and 0.45.  This value agrees well with our analysis in Sect.~\ref{sp_class}, where the best fit
to the observed continuum was obtained with an M3 giant spectrum reddened
with $E_{B-V} \simeq 0.35$.

However, as proposed in \citet{kst10} and confirmed by spectropolarimetric
observations in \citet{kt13}, the observed stellar continuum in V4332~Sgr is
not a direct spectrum of the main, central object, but results from scattering
the central-star radiation on dust grains. Since scattering on cosmic dust 
grains usually causes the incident spectrum to be bluer, the result of $E_{B-V} = 0.35$ 
obtained in Sect.~\ref{sp_class}, as well as that in \citet{kst10}, may be
regarded as a lower limit to the reddening of V4332~Sgr.

\citet{kst10} proposed that the emission features observed in the
spectrum of V4332~Sgr result from resonant scattering of the central-star
radiation by molecules and atoms in a circumstellar matter.
They also concluded that the strongest
emission lines, such as \NaI\,5890,5896\,\AA\ and 
\KI\,7665,7699\,\AA,
are optically thick in the sense that they scatter all the incident stellar
radiation. The main argument was that the observed intensity ratio of 
the two components of the \NaI\ and \KI\ doublets was close to 1:1, while in
the optically thin case this ratio should be 2:1. The spectrum of V4332~Sgr
we discuss here shows much more numerous and strong emission
features than that of \citet{kst10}. In addition to the \NaI\ and \KI\ 
doublets,
we therefore searched for other emission lines, which are likely to be
optically thick. These are the \CrI\ lines at 4254,4275,4290\,\AA. 
In the optically thin case, the intensity ratio
of these lines should be 1.8:1.4:1.0. As can be seen from
Table~\ref{atom_tab}, the observed ratio is 1.02:0.89:1.0. Almost certainly,
the \CaI\,4227\,\AA\ line is also optically thick because it presents
the highest value of the product of element abundance and 
transition probability of all the observed emission features in our
spectrum. Finally, the line ratio of the \MnI\ lines at 
4031,4033,4034\,\AA\ suggests that these lines
are almost optically thick, at least the first one. 
The theoretical optically thin line ratio for these \MnI\ lines
is 2.0:1.5:1.0. The lines are strongly blended, but from the observed profile
we can estimate a ratio of 1.4:0.8:1.0.

\begin{figure}
  \includegraphics[width=\columnwidth]{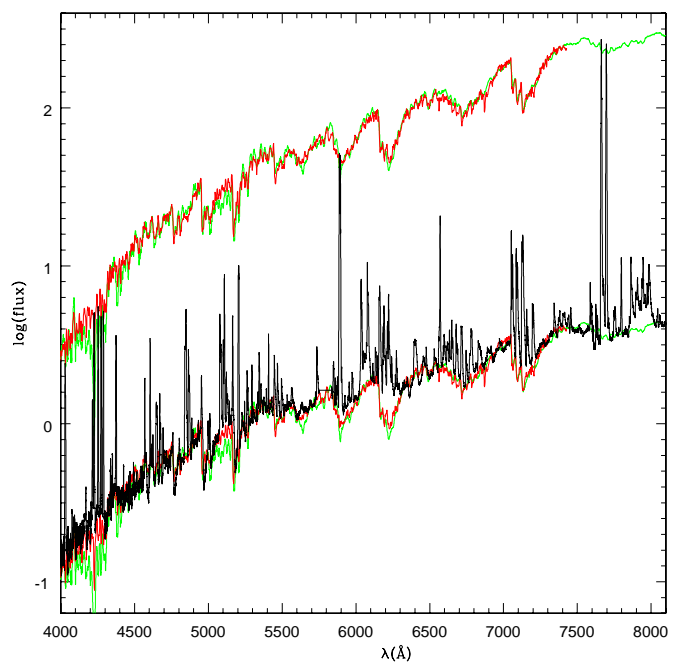}
  \caption{Reference (standard and model giant) spectra compared to the observed 
  spectrum of V4332~Sgr (black). Red: M3 standard spectrum, 
  green: model atmosphere ($T_{\rm eff} = 3600$~K, 
   log~$g$ = 0.5). Reference spectra in the lower plots are reddened with 
  $E_{B-V} = 0.35$. Reference spectra in the upper plots are reddened with
  $E_{B-V} = 0.75$ and shifted upward
  to match the highest fluxes of the strongest
  emission lines.}
\label{sp_red_fig}
\end{figure}    

In this way, we have a set of emission lines 
whose monochromatic fluxes in the central, most opaque regions of
their profiles are expected to measure the monochromatic flux from 
the central star. These
lines span a wide spectral region,  4000--8000~\AA, which means
that they can be used
to estimate the interstellar reddening when compared to a reference spectrum
that is assumed to represent the spectrum of the central object of V4332~Sgr.
This is done in Fig.~\ref{sp_red_fig}. The lower part of the figure presents
essentially the same as Fig.~\ref{sp_class_2_fig} (i.e. the standard M3 and
model, $T_{\rm eff} = 3600$~K, spectra reddened with $E_{B-V} = 0.35$ 
($R \equiv A_{\rm V}/E_{B-V} = 3.1$ was used throughout this paper) and
compared to the observed spectrum) but in the logarithm scale of the flux. The
upper plots show the same reference spectra, but shifted upwards to match the
highest fluxes of the strongest, that is, optically thick, emission lines,
namely \MnI\,4031\,\AA, \CaI\,4227\,\AA,
\CrI\,4254,4275,4290\,\AA,
\NaI\,5890,5896\,\AA, and 
\KI\,7665,7699\,\AA. A pure upward shift did not suffice
to fit the level of the highest fluxes in the lines. It was also necessary to
increase the reddening to $E_{B-V} \simeq 0.75$. Thus we can conclude that
the emission-line spectrum of V4332~Sgr is significantly more reddened than the
stellar-like continuum. As discussed above, the difference in the reddening
is most likely due to the bluering of the stellar continuum that is scattered on 
circumstellar dust grains.

There is a process, however, that can affect the relative fluxes of the
lines we have used to estimate the reddening of the emission spectrum. All
the lines we have considered above are from resonant transitions in atoms.
If the lines are optically thick, the resonance photons must suffer from numerous
scattering before they escape the emitting region. This increases the chances of
being absorbed by dust in the region. Since the absorption coefficent of dust
grains is expected to increase with decreasing wavelength, the lines at
4000--4300\,\AA would suffer more from dust absorption than the
lines at 7600--7800\,\AA. Detailed line transfer calculations are
neccesary to investigate this effect, which is beyond the scope of the
present study. The process, if effective, would mimic an
additional interstellar extinction. Therefore we conclude from
the discussion in
this section that the
interstellar reddening to V4332 is $0.35 \la E_{B-V} \la 0.75$.

\section{Radial velocity  \label{rad_vel}}

All clear absorption features in the spectrum of V4332~Sgr can be identified
as molecular bands (mostly TiO). They are wide, shallow, and usually
filled near their heads with strong emission components. We were not able
to identify any atomic absorption line. This precludes a reliable determination of
the radial velocity of V4332~Sgr on the basis of its
absorption spectrum.

However, many of the observed atomic emission lines
are relatively narrow (FWHM $\la$1~\AA, see Col. 5 in Table~\ref{atom_tab}), 
and they can be used 
to determine the radial velocity of the object. The strongest 
emission lines are not particularly suitable for this purpose
because they are wide and their observed wavelengths are poorly 
determined. We have selected a set of 28 unblended
lines of medium intensities. All of them gave consistent values of the radial
velocity, ranging between $-$83 and $-$68~km\,s$^{-1}$. The mean value and
standard deviation derived from this sample are 
$-75.1 \pm 3.1$~km\,s$^{-1}$.
This value can be compared with $-65 \pm 7$~km\,s$^{-1}$ obtained in
\citet{kst10}, but that was an estimate based on strong and wide emission
lines. 

We note that these estimates refer to the line-emitting region, and it is not clear to what extent they measure the radial
velocity of the main, stellar-like object of V4332~Sgr.
If these values measure the real radial velocity of V4332~Sgr, then the
object does not follow the standard rotation of the Galaxy 
\citep[see e.g.][]{blitz} because the heliocentric radial velocity at any distance is
expected to be $\ga -10$~km\,s$^{-1}$ for the position of V4332
Sgr.

\section{Interstellar lines and distance to V4332~Sgr \label{distance}}

There is no reliable estimate of the distance to V4332~Sgr. \citet{martini}
proposed 300~pc assuming that the object was a K-type giant at maximum of
the 1994 outburst. \citet{tcgs05} obtained 1.8~kpc assuming that the
progenitor was a solar-type main-sequence star. \citet{kimes} estimated 2.9
or 10~kpc depending on whether luminosity class V or III was
assumed for the
progenitor. \citet{kst10} derived a kinematic distance $\ga$1.0~kpc from
the radial velocity of the interstellar absorption lines of the 
\NaI\,5890,\,5896\,\AA,\ and \KI\,7665,\,7699\,\AA\ lines. These authors
identified a single component of these lines in their spectrum of V4332~Sgr.
Our spectrum is of a much better quality than that of \citet{kst10}, and we
can identify several components of the interstellar \NaI\ lines.
 
\begin{figure}
  \includegraphics[width=\columnwidth]{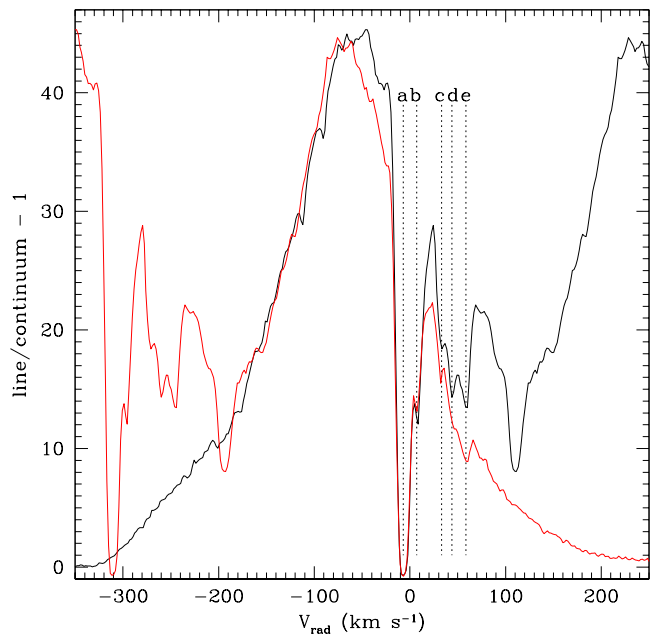}
  \caption{Profiles of the \NaI\,5890\,\AA\ (black) and 5896\,\AA\ (red)
lines superimposed on the radial velocity scale. Components of the
interstellar absorptions are marked with vertical dotted lines.}
\label{NaI_fig}
\end{figure}    

\begin{table}
\begin{center}
\caption{Components of the interstellar absorption in the profile of 
the \NaI\,5890\,\AA\ line (see Fig.~\ref{NaI_fig}).}
\label{NaI_tab}
\begin{tabular}{ c c c c }
\hline
 designation  & $\lambda_{\rm obs}$(\AA)  & $V_{\rm r}$(km\,s$^{-1}$) &
distance (kpc)
\\
\hline
  a  &  5889.779 &  $-$8.7 & 0.48 \\
  b  &  5890.104 &  ~~7.8 &  2.50 \\
  c  &  5890.573 &  31.7 &  4.23 \\
  d  &  5890.817 &  44.2 &  4.85 \\
  e  &  5891.110 &  59.1 &  5.44 \\
\hline
\end{tabular}
\end{center} 
\end{table}

Figure~\ref{NaI_fig} displays the observed profiles of the \NaI\,5890 and 5896\,\AA\
lines superimposed on each other, plotted on the radial velocity scale. 
Up to five components of the 
interstellar absorptions
are clearly seen, particularly in the \NaI\,5890\,\AA\ line.
They are marked with letters in the
figure and are listed in Table~\ref{NaI_tab}. The observed wavelengths of the
individual components were
obtained by fitting Gaussian profiles to the observed features. 
Note that the absorption feature seen at 
$V_r \simeq 110$~km\,s$^{-1}$ in the \NaI\,5890\,\AA\ line, or at
$V_r \simeq -195$~km\,s$^{-1}$ in the \NaI\,5896\,\AA\ line, cannot be of
interstellar origin because it has no counterpart in the other line.

The observed wavelengths of the interstellar components given in the second
column of Table~\ref{NaI_tab} can be used to calculate heliocentric radial
velocities, which are listed in the third column of the table. These values,
after being transformed into radial velocities in the LSR frame (adding
12.0~km\,s$^{-1}$ for the position of V4332~Sgr) and 
adopting the Galactic rotation curve of \citet{blitz}, give
estimates of distances of the interstellar regions
that are responsible for the observed features. The results are listed in
the last column of Table~\ref{NaI_tab}. From these data we can conclude that
V4332~Sgr is located at a distance $\ga$5.5~kpc. This together
with the Galactic latitude of
the object ($b = -9\fdg40$) implies that V4332~Sgr is
situated $\ga$0.89~kpc below the Galactic plane.

\section{Evolution of the spectrum of V4332~Sgr between 2005 and 2009
\label{spec_evol}}

\begin{figure}
  \includegraphics[width=\columnwidth]{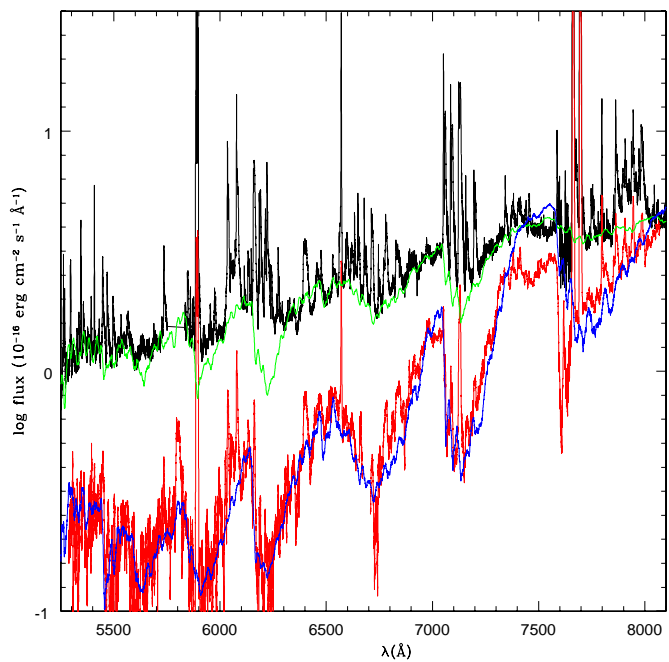}
  \caption{Comparison of the spectrum of V4332~Sgr in 2005 (black: this
paper) to that observed in 2009 (red: \citet{kst10}).
  Green: model atmosphere ($T_{\rm eff} = 3600$~K, 
   log~$g$ = 0.5) reddened with 
  $E_{B-V} = 0.35$ and fitted to the 2009 spectrum. Blue: model
atmosphere ($T_{\rm eff}= 3300$~K, log~$g$ = 0.0) reddened with $E_{B-V} =
0.35$ and fitted to the 2005 spectrum.}
\label{sp_evol_fig}
\end{figure}

Four years after the spectrum described in this paper had been registered, 
a high-resolution spectrum of V4332~Sgr was obtained by \citet{kst10} using
the Subaru telescope. As can be seen from Fig.~\ref{lc_fig}, the object
in meantime faded by $\sim$2~mag. in the $V$ band.
Figure~\ref{sp_evol_fig} directly compares the spectra obtained in these two
epochs. Two MARCS model spectra fitted to the observed spectra are also plotted
in the figure. Note that both model spectra were reddened with $E_{B-V} =
0.35$. Note also that the model fit to the 2009 spectrum is not perfect in
the long-wavelength range. This point, a possible source of extra absorption
in the 7300--7500~\AA\ range in particular, was discussed in \citet{kst10}.

One of the clear differences between the 2005 and 2009 spectra displayed in
Fig.~\ref{sp_evol_fig} is that the 2009 stellar-like continuum is definitely of 
a later spectral type than that of 2005. Indeed, \citet{kst10} classified the
2009 spectrum as M5-6, while we have concluded that it is an M3 type in
Sect.~\ref{sp_class}. This has obvious consequences on the effective
temperature of the central stellar-like object in V4332~Sgr. Our MARCS model
fitted to the observed spectra has $T_{\rm eff} \simeq 3300$~K in 2009 compared to
$T_{\rm eff} \simeq 3600$~K in 2005. However, when fitting the model spectra
to the observed spectra in the flux scale, we had to assume that the emitting
surface in 2009 increased by a factor of $\sim$2.08 comparing to 2005. This
implies an increase of the effective radius of the object by a factor of
$\sim$1.44 if a spherically symmetric case is assumed. The luminosity of the
object would thus have increased by a factor of $\sim$1.47 between 2005 and 2009.

\begin{figure}
  \includegraphics[width=\columnwidth]{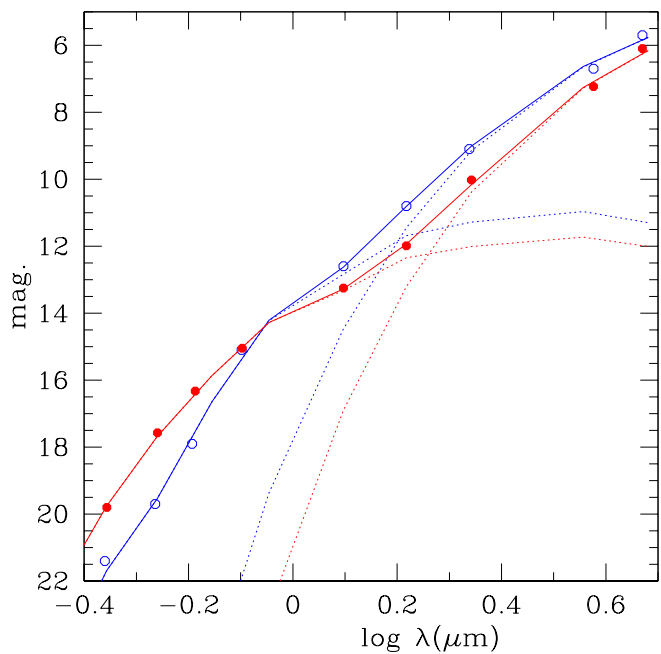}
  \caption{Comparison of the photometric measurements of V4332~Sgr in 2003 
(red filled symbols: \citet{tcgs05})
to those obtained in 2009 (blue open symbols: \citet{kst10}).
Full curves: model photometric spectra obtained by summing a standard stellar
component with a black-body dust component (individual components are shown with
dotted curves) reddened with  $E_{B-V} = 0.35$. See text for more details of
the modelling.}
\label{phot_evol_fig}
\end{figure}    

Similar conclusions to these from the optical spectroscopy of V4332~Sgr 
can be derived from analysing photometric measurements of the object in 
the optical ($BVRI$) and near-IR ($JHKLM$) bands. In
Fig.~\ref{phot_evol_fig} we plot the results of photometric
measurements made in 2003 and compiled in \citet{tcgs05} (red filled
symbols), as well as those obtained in 2009 and given in \citet{kst10} (blue
open symbols). (There were no near-IR measurements of V4332~Sgr in 2005,
but, as can be seen from Fig.~\ref{lc_fig}, the object did not significantly
evolve photometrically between 2003 and 2005.) The data were fitted with
standard stellar photometric spectra supplemented with black-body dust
components in the same way as in \citet[][for details of the fitting
procedure see \citet{tyl05}]{tcgs05}. The final fits (sum of the stellar and
dust components reddened with $E_{B-V} = 0.35$) are shown with the full curves 
in Fig.~\ref{phot_evol_fig}.
The fit for the 2003 data (red in the figure) is the same as those in 
\citet[][see their Fig.~5]{tcgs05}, that is, an M2.7 supergiant 
\citep[$T_{\rm eff} = 3620$~K according to the temperature scale of][]{levesque} 
and a 750~K black-body dust. The fit for 2009 (blue in the figure) consists of 
an M6.2 ($T_{\rm eff} = 3280$~K) supergiant and a 900~K black-body dust.
When the luminosities of the components are compared, it appears that both 
components increased by a similar factor between 2003 and 2009, 
that is, the stellar component by 1.48 times, the dust component
by 1.51 times.

Thus we can quite safely conclude that the optical decline of V4332~Sgr observed 
between 2005 and 2007 (see Fig.~\ref{lc_fig}) was due to a decrease of 
the effective temperature of the central stellar-like object by 300--350~K. 
The object expanded during this event, however, and its luminosity 
increased by $\sim$50\%.

Figure~\ref{sp_evol_fig} reveals another important difference between the
spectra in 2005 and 2009. In the 2009 spectrum the emission features are
significantly fainter than in 2005, not only in absolute flux scale, but also
relative to the M-type continuum. The fading scale is not the
same for all the features, however. The strongest atomic lines, for instance those of 
\NaI\ and \KI, show a similar strength relative to the local continuum in
both epochs. On the other hand, many fainter features that were
clearly present in 2005
completely disappeared in 2009. To investigate the problem more
quantitatively, we calculated relative contributions of the emission 
features to the total flux in selected wavelength ranges in both spectra.
For a given wavelength range, we derived $F_{\rm obs}$ as an integral of 
the observed flux over the wavelength and $F_{\rm mod}$ as an integral of the
MARCS model flux. We assumed that the MARCS spectrum, when fitted to the
observed spectrum as in Fig.~\ref{sp_evol_fig},
is a good representation of the stellar-like continuum of V4332~Sgr. Then
we can define a parameter $f_{\rm em} = (F_{\rm obs}-F_{\rm mod})/F_{\rm obs}$
as a measure of the relative contribution of the emission features to the
total observed flux.

For the whole spectral range common for both spectra, that is,
for 5500--8000~\AA,
we derive $f_{\rm em} = 0.43$  for the 2005 VLT spectrum and  0.25 for the 2009
Subaru spectrum. Thus the global contribution of the emission
spectrum to the total observed flux decreased by $\sim$40\% between 2005 and
2009. For narrow spectral ranges encompassing the \NaI\ and \KI\ emission
lines, that is, for 5880--5910~\AA\ and 7650--7720~\AA, the result is $f_{\rm em} =
0.93$ and 0.91 in 2005 compared to $f_{\rm em} = 0.91$ and
0.89 in 2009. Thus, as stated above, the strongest emission features remain
practically at the same level when compared to the local continuum.
For the 6560--6580~\AA\ range, which includes the \CaI\ line, the
figures are $f_{\rm em} = 0.60$ and 0.49 for 2005 and 2009,
which is a decrease of 18\%. More important, fading affected molecular 
emission features. For series of the TiO$\gamma$ bands observed
within 6590--6760~\AA\ and 7040--7220~\AA\ we derive $f_{\rm em} = 0.27$ and
0.57 in 2005, while for the 2009 spectrum the figures are
0.14 and 0.30. In other words, these emission features faded by a factor of
2 relative to the continuum between 2005 and 2009. 
Even greater fading affected the VO-B-X(0-0) bands
gathered between 7850--8010~\AA. The result is $f_{\rm em} = 0.37$ and 0.13
(2005 versus 2009) in this case. Thus a general (although not strict) tendency 
is that the fainter emission feature, the greater the fading effect.

This behaviour can be easily explained within the scenario proposed by
\citet{kst10}, according to which the emission features in V4332~Sgr are
produced by radiative excitation of atomic and molecular resonant transitions
in a circumstellar matter by strong radition from the
central stellar-like object, which is invisible to us. The fading of the emission features can then be
understood in terms of a decreasing optical thickness of the circumstellar
matter, for example due to its expansion. For optically thin transitions the fading
would be proportional to the decrease of the optical thickness. However, for
optically thick lines, as is probably the case of the \NaI\ and \KI\ resonant transitions, a modest decrease of their optical thickness would not affect their
observed strengths relative to the continuum. The flux in these lines is 
limited by the available radiative flux from the central source. 
Thus these emission features fade proportionally to the incident flux, 
so that their ratio to the observed M-type continum remains unchanged.

\section{Summary and discussion  \label{discussion}}

We have reduced the best-quality optical spectrum ever
obtained for V4332~Sgr. The spectrum was recorded in April/May~2005 with the
VLT/UVES equipment and is unique not only because of its quality, but also 
because the object has considerably evolved since that time. Most of
the emission features that were numerous in
2005 disappeared, which means that it will not be
possible, in the future, to repeat the observations of the object in 
a similar stage. 
Therefore we decided to reduce and publish the spectrum, 
so that it is made
available to the astrophysical community. We presented the
spectrum and results of its measurements and general analysis. 
Forthcoming papers will be
devoted to detailed analyses of the emission features, their profiles,
intensities, and chemical species that produce them.

The spectrum is dominated by numerous atomic and molecular emission features
that are superimposed on an M-type continuous spectrum. Among the emission features
we have identified over 70
atomic lines belonging to 11 elements (Na, Mg, Al, K, Ca, Cr, Mn, Fe, Rb,
Sr, and Ba) (see Table~\ref{atom_tab}) and about 140 bands belonging to 6 molecules
(AlO, ScO, TiO, VO, CrO, and YO) (see Table~\ref{molec_tab}).
There is no other late-type stellar object with
an emission spectrum that rich and intense.

The underlying stellar-like continuum in the spectrum of V4332~Sgr can be
classified as $\sim$M3 (see Sect.~\ref{sp_class}). 
A giant spectrum of this spectral type fits the observed 
continuum quite well. Moreover, a MARCS model spectrum calculated for an
effective temeprature characteristic of $\sim$M3, that is, $\sim$3600~K, 
reproduced the observations. 
Strong molecular emissions partly filling the absorption
features in the observed continuum did not allow us to analyse the stellar-like continuum of V4332~Sgr in more detail.

From fitting the standard stellar spectrum and the MARCS model
spectrum to the observed continuum of V4332~Sgr, we derived an
interstellar reddening, $E_{B-V} \simeq 0.35$ (see Sect.~\ref{reddening}). This
value is consistent with previous determinations \citep{tcgs05,kimes,kst10}.
However, since the observed continuum is assumed to 
result from scattering on dust grains,
this value probably
underestimates the interstellar reddening. Therefore we
attempted to estimate the extinction from the strongest
emission lines because the highest flux in these lines is expected to follow 
the flux from the central star. Comparing these data with the
standard M3 spectrum and the MARCS model, we obtained $E_{B-V} \simeq 0.75$. This
value should be regarded as an upper limit to the real reddening because of
possible dust absorption effects of
multiply scattered resonant-line photons.
Thus we concluded that the interstellar reddening of 
V4332~Sgr is $0.35 \la E_{B-V} \la 0.75$.

In Sect.~\ref{rad_vel} we derived the radial velocity of V4332~Sgr from
a set of relatively narrow medium-intensity emission lines. Our result of
$-75.4 \pm 3.5$~km\,s$^{-1}$ is consistent with the results from other
determinations \citep{martini,tcgs05,kst10} in the sense that all of them
gave negative values, although the values range between $-$180 to
$-$56~km\,s$^{-1}$. As already discussed in \citet{tcgs05}, these values show
that V4332~Sgr does not follow the Galactic rotation curve because from the
latter one would expect a heliocentric 
radial velocity of $\ga -10$~km\,s$^{-1}$ for any distance for
the object coordinates.
The problem becomes even more evident if one takes into account the lower
limit of the distance derived in Sect.~\ref{distance}, which, assuming the
Galactic rotation curve, implies a radial velocity $\ga 60$~km\,s$^{-1}$.
One possible explanation is that what we measure in the emission lines is
not the radial velocity of the object, but
the expansion of the matter ejected in 1994. It is difficult to
understand, however, why a decade after the eruption, we would not see the receding
part of the ejecta, especially because the radial velocity was determined from
medium-intensity lines, which are expected to be optically thin. Another
possibility is that V4332~Sgr is not a Galactic disc object, hence its
strange radial velocity. This interpretation would be supported by the
relatively large distance of the object from the Galactic plane of 
$\ga$0.89~kpc (see Sect.~\ref{distance}).

Our comparison of the 2005 VLT spectrum to the spectrum obtained in 2009 that was
described in \citet{kst10} (see Sect.~\ref{spec_evol}) showed that the M-type
continuum evolved from an $\sim$M3 spectral type to M5-6, so that the
effective temperature of the central stellar object decreased by 300-350~K.
The same conclusion also results from photometric measurements made in 2003
and 2009. This relatively small ($\sim$10\,\%) decrease in the effective
temperature is fully responsible for the optical fading of the object
observed in 2006. The object expanded during this event,
however, so that its
luminosity increased by $\sim$50\,\%. This is evident not only from fitting
the optical observations, but also from the infrared photometry, which is of
particular importance because the infrared emission dominates the observed
spectral energy distribution of the object \citep[see e.g.][]{kst10}.
The origin of this behaviour is not clear, but it might be a manifestation of
a long-term relaxation of the remnant of the presumable merger event in
1994. The main object is embedded in a massive dusty envelope, probably in
the form of a thick disc. An accretion event from the envelope to the central
stellar object can be invoked here.

The good resolution and quality of the spectrum we analysed
allowed us to detect several components of the interstellar absorption in
the \NaI\,5890 and 5896\,\AA\ lines (see Sect.~\ref{distance}). 
The radial velocities of these features, if interpreted as due to interstellar
matter moving according to the Galactic rotation curve, imply a lower limit
to the distance of V4332~Sgr of $\sim$5.5~kpc. This result has important
consequences on estimates of the global parameters of V4332~Sgr. 
In particular, the luminosity
of the object during the 1994 eruption, as obtained from photometric
measurements in \citet{tcgs05}, increases now to $\ga 4.5 \times
10^4$~L$_\odot$, while its effective radius becomes $\ga 450$~R$_\odot$.
These values are not as high as those derived for V838~Mon in its 2002
eruption \citep[see e.g.][]{tyl05}, but are of a similar order as those of
V1309~Sco in its 2008 eruption \citep{thk11}.

There are, in fact, other similarities between V4332~Sgr and V1309~Sco. The
eruptions of both objects were of a similar time scale, that
is, about one
month. That of V4332~Sgr was probably slightly longer because its rising part was
not observed. The progenitor of V4332~Sgr was variable, and the archive data
suggest that it was slowly rising in brightness on a time scale of decades 
\citep{kimes07,goran07}. This resembles the slow systematic rise of V1309~Sco 
observed during a few years before its eruption \citep{thk11}. 
The remnants of both objects
are strongly dominated by infrared dust emission \citep{kst10,nicholls}.
Finally, an optical and near-IR spectrum of V1309~Sco obtained in 2012
\citep{kmts15} shows an emission-line spectrum similar to that of
V4332~Sgr, although not as rich and intense as in the latter case. Clearly,
the remnant of V1309~Sco is heavily embedded in dust as was the case of
V4332~Sgr. It is therefore tempting to suggest that the nature of the progenitor 
and the eruption of V4332~Sgr was similar to those of V1309~Sco.
We know that the eruption of V1309~Sco resulted from merger of a contact
binary \citep{thk11}.

\begin{acknowledgements} 
We are grateful to Dipankar P. K. Banerjee, PI of the ESO-VLT 075.D-0511 
observing programme, for the spectroscopic observations of V4332~Sgr made in
2005, on which this paper is based. Many thanks to the referee for the
comments on an earlier version of the paper.
\end{acknowledgements}

\clearpage

\setcounter{table}{1}
\begin{table*}
\caption{
  Atomic lines of the spectrum of V4332 Sgr. The flux is 
listed in 10$^{-16}$ erg\,cm$^{-2}$\,s$^{-1}$, the wavelengths and FWHM in \AA.
}
\label{atom_tab}
\begin{center}
\begin{tabular}{     l l l r l l}
\hline
 ${\lambda}_{\rm obs}$ &
 ${\lambda}_{\rm lab}$ &  
 identification    & 
 flux~~~~~~~~~~~~~ &
 FWHM              &
 notes             \\
\hline

  3860.47 & 3859.91 & FeI                   &     0.37  ($\pm$0.41)  &  2.1: & f \\
 
  3885.48 & 3886.28 & FeI                   &     2.12  ($\pm$0.48) &  2.55  &  \\

  3922.26 & 3922.91 & FeI                   &     1.38  ($\pm$0.35) &  2.5:  &   \\

  3929.29 & 3930.30 & FeI                   &     0.89  ($\pm$0.30) &  2.1:  &   \\

  3932.86 & 3933.66 & CaII                  &     0.77  ($\pm$0.49) &  3.4:  &   \\

  3943.40 & 3944.01 & AlI                   &     1.07  ($\pm$0.33) &  2.3:  &  \\

  3960.79 & 3961.52 & AlI                   &     6.64  ($\pm$0.38) &  2.70  &  \\

  3967.84 & 3968.47 & CaII                  &     1.73  ($\pm$0.50) &  3.5:  &  \\

  4029.8~ & 4030.75 & MnI                   &    11.60  ($\pm$1.08) &  5.3:  & $\rceil$B\\
  4031.8~ & 4033.06 & MnI                   &                       &        &  \,|B   \\
  4033.5~ & 4034.48 & MnI                   &                       &        & $\rfloor$B   \\

  4076.66 &          & ?                     &     0.20  ($\pm$0.08) &  0.6:  & f \\

  4170.99 &          & ?                     &     0.58  ($\pm$0.32) &  2.5:  & f \\

  4205.66 & 4206.70 & FeI                   &     0.20  ($\pm$0.11) &  0.9:  & f \\

  4215.16 & 4216.18 & FeI                   &     4.39  ($\pm$0.41) &  3.17  & \\

  4226.9~ & 4226.73 & CaI                   &    20.17  ($\pm$0.72) &  3.31  & \\

  4253.58 & 4254.33 & CrI                   &    18.43  ($\pm$0.39) &  3.00  &  \\

  4274.01 & 4274.80 & CrI                   &    16.02  ($\pm$0.39) &  3.06  &  \\

  4289.08 & 4289.72 & CrI                   &    18.04  ($\pm$0.41) &  3.17  &  \\

  4336.46 & 4337.57 & CrI                   &     0.28  ($\pm$0.08) &  0.58  &  \\

  4338.44 & 4339.45 & CrI                   &     0.62  ($\pm$0.13) &  0.98  & $\rceil$B  \\
           & 4339.72 & CrI                   &                       &        & $\rfloor$B   \\

  4343.39 & 4344.50 & CrI                   &     0.34  ($\pm$0.09) &  0.66  & \\

  4349.92 & 4351.05 & CrI                   &     0.39  ($\pm$0.11) &  0.87  & b \\

  4350.79 & 4351.76 & CrI                   &     0.30  ($\pm$0.08) &  0.60  & b \\

           & 4359.63 & CrI                   &                       &        & p \\

  4370.22 & 4371.26 & CrI                   &     0.20  ($\pm$0.09) &  0.69  &  \\

  4374.94 & 4375.93 & FeI                   &    12.19  ($\pm$0.84) &  2.43  & \\

           & 4384.98 & CrI                   &                       &         & p \\

  4426.23 & 4427.31 & FeI                   &     0.47  ($\pm$0.07) &  0.58  & \\

  4460.53 & 4461.65 & FeI                   &     0.32  ($\pm$0.11) &  0.93  & \\

  4495.73 & 4496.84 & CrI                   &     0.11  ($\pm$0.07) &  0.60  & \\

  4544.72 & 4545.94 & CrI                   &     0.25  ($\pm$0.15) &  1.2:  & \\

  4552.95 & 4554.12 & BaII                  &     0.23  ($\pm$0.07) &  0.59  & \\

  4569.95 & 4571.10 & MgI                   &     4.03  ($\pm$0.12) &  0.95  & \\

  4578.87 & 4580.04 & CrI                   &     0.16  ($\pm$0.08) &  0.67  & \\

  4599.62 & 4600.74 & CrI                   &     0.44  ($\pm$0.16) &  1.29  & \\

  4606.15 & 4607.33 & SrI                   &     8.85  ($\pm$0.48) &  1.33  & \\

  4612.29 & 4613.36 & CrI                   &     0.20  ($\pm$0.12) &  0.92  & \\

  4614.98 & 4616.12 & CrI                   &     0.39  ($\pm$0.12) &  0.92  & \\

  4624.98 & 4626.17 & CrI                   &     0.65  ($\pm$0.13) &  1.04  & \\

  4644.99 & 4646.15 & CrI                   &     0.86  ($\pm$0.12) &  0.98  & \\

  4650.18 & 4651.28 & CrI                   &     0.13  ($\pm$0.06) &  0.50  & b,c \\

  4650.98 & 4652.15 & CrI                   &     0.25  ($\pm$0.07) &  0.56  & b,c \\

  4932.87 & 4934.08 & BaII                  &     0.33  ($\pm$0.06) &  0.59  & \\

  5109.14 & 5110.41 & FeI                   &    28.39  ($\pm$1.12) &   1.48  & b \\

           & 5166.28 & FeI                   &                       &        &  B,u \\
           
           & 5168.90 & FeI                   &                       &        &  B,u\\

  5203.07 & 5204.51 & CrI                   &    13.93  ($\pm$0.20) &  1.29: & b \\

  5204.68 & 5206.04 & CrI                   &    16.17  ($\pm$0.19) &  1.22: & b \\

  5207.04 & 5208.42 & CrI                   &    24.17  ($\pm$0.21) &  1.37: & b \\

  5246.23 & 5247.57 & CrI                   &     1.15  ($\pm$0.14) &  0.91  & \\

  5262.86 & 5264.15 & CrI                   &     2.46  ($\pm$0.12) &  0.79  & b,c\\

  5264.49 & 5265.72 & CrI                   &     0.30  ($\pm$0.05) &  0.3:  & b,c\\

  5268.27 & 5269.54 & FeI                   &     1.05  ($\pm$0.10) &  0.61  & c \\

  5295.34 & 5296.69 & CrI                   &     2.53  ($\pm$0.15) &  0.97  & b \\

  5296.94 & 5298.28 & CrI                   &     2.31  ($\pm$0.12) &  0.77  & b \\

  5299.18 & 5300.74 & CrI                   &     0.73  ($\pm$0.26) &  1.5:  & b,f \\

\hline
\end{tabular}
\end{center}
\end{table*}

\setcounter{table}{1}
\begin{table*}
\caption{  Continued}
\begin{center}
\begin{tabular}{     l l l r l l}
\hline
 ${\lambda}_{\rm obs}$ &
 ${\lambda}_{\rm lab}$ &  
 identification    & 
 flux~~~~~~~~~~~~~ &
 FWHM              &
 notes             \\
\hline

  5326.72 & 5328.04 & FeI                   &     0.99  ($\pm$0.24) &  1.4:   & s \\

  5344.46 & 5345.80 & CrI                   &     4.54  ($\pm$0.16) &  0.93  & b \\

  5346.94 & 5348.31 & CrI                   &     2.93  ($\pm$0.18) &  1.0:  & b \\

           & 5394.68 & MnI                   &                       &        & p,B,u \\

  5408.5~ & 5409.77 & CrI                   &     6.23  ($\pm$0.22) &  0.82  & \\

  5431.17 & 5432.55 & MnI                   &     0.58  ($\pm$0.14) &  0.84  & f,c \\

  5534.20 & 5535.48 & BaI                   &     0.96  ($\pm$0.22) &  1.4:  & \\

  5666.76 & 5668.28  & MnI                   &     0.08  ($\pm$0.05) &  0.3:  & f,c \\

  5688.96 & 5690.43  & MnI                   &     0.21  ($\pm$0.10) &  0.65  & f,c \\

  5727.10 & 5728.57  & MnI                   &     0.43  ($\pm$0.26) &  1.6:  & \\

  5861.20 & 5862.69  & MnI                   &     0.99  ($\pm$0.19) &  1.27  & c\\

  5889.2~ & 5889.95 & NaI                   &   252.10   ($\pm$0.60) &  4.0: E,b \\

  5894.6~ & 5895.92 & NaI                   &   183.35   ($\pm$0.40) &  2.8: E,b \\

  6279.10 & 6280.62 & FeI                   &     0.24  ($\pm$0.06) &  0.4: & f,b \\

  6328.26 & 6330.09 & CrI                   &     0.41  ($\pm$0.18) &  1.3: & s \\

  6571.10 & 6572.78 & CaI                   &    49.19  ($\pm$0.17) &  1.12  & B \\

  7663.~~ & 7664.91 & KI                    &  1358.74  ($\pm$2.26) &  3.34  & E,b \\

  7697.~~ & 7698.97 & KI                    &  1291.33  ($\pm$2.19) &  3.23  & E,b \\

  7798.52 & 7800.27 & RbI                   &    37.01  ($\pm$2.10) &  2.11  & b \\

  7946.14 & 7947.60 & RbI                   &     6.15  ($\pm$0.49) &  1.55 & B\\
\hline
\end{tabular}
\end{center}
\end{table*} 


\setcounter{table}{2}
\begin{table*}
\caption{
  Molecular bands of the spectrum of V4332 Sgr. The flux is listed in 10$^{-16}$ erg\,cm$^{-2}$\,s$^{-1}$ 
, the wavelengths in \AA.}
\label{molec_tab}
\begin{center}
\begin{tabular}{ l l r  l}
\hline
 ${\lambda}_{\rm lab}$ &  
 identification    & 
 flux~~~~~~~~~~~~~ &
 notes             \\
\hline
            4470.5 & AlO~~B$^{2}\Sigma^{+}$-X$^{2}\Sigma^{+}$~~(2,0) &     0.38  ($\pm$0.06) & \\

            4494.0 & AlO~~B$^{2}\Sigma^{+}$-X$^{2}\Sigma^{+}$~~(3,1) &     0.28  ($\pm$0.05) & f \\

            4516.4 & AlO~~B$^{2}\Sigma^{+}$-X$^{2}\Sigma^{+}$~~(4,2) &     0.29  ($\pm$0.08) & f \\

            4648.2 & AlO~~B$^{2}\Sigma^{+}$-X$^{2}\Sigma^{+}$~~(1,0) &     9.07  ($\pm$0.61) & b \\

            4672.0 & AlO~~B$^{2}\Sigma^{+}$-X$^{2}\Sigma^{+}$~~(2,1) &     6.43  ($\pm$0.55) &  \\

            4694.6 & AlO~~B$^{2}\Sigma^{+}$-X$^{2}\Sigma^{+}$~~(3,2) &     2.26  ($\pm$0.60) & \\

            4715.5 & AlO~~B$^{2}\Sigma^{+}$-X$^{2}\Sigma^{+}$~~(4,3) &     0.62  ($\pm$0.42) & f \\

            4735.8 & AlO~~B$^{2}\Sigma^{+}$-X$^{2}\Sigma^{+}$~~(5,4) &                       & p \\

            4760.9 & TiO~~$\alpha$~~(2,0)~~R$_2$ &     0.41  ($\pm$0.09) & c \\

            4804.3 & TiO~~$\alpha$~~(3,1)~~R$_2$ &                       &  p \\

            4842.3 & AlO~~B$^{2}\Sigma^{+}$-X$^{2}\Sigma^{+}$~~(0,0)          &    43.52  ($\pm$0.83) & \\

            4866.4 & AlO~~B$^{2}\Sigma^{+}$-X$^{2}\Sigma^{+}$~~(1,1)          &    16.26  ($\pm$0.67) & \\

            4889.0 & AlO~~B$^{2}\Sigma^{+}$-X$^{2}\Sigma^{+}$~~(2,2)          &     3.63  ($\pm$0.89) & c \\

            4954.6 & TiO~~$\alpha$~~(1,0)     &     5.02  ($\pm$0.51) & c \\

            4999.1 & TiO~~$\alpha$~~(2,1)     &     2.82  ($\pm$0.74) & c \\

            5010.5 & VO~~C\,$^{4}\Sigma^{-}$-X\,$^{4}\Sigma^{-}$~~(3,0) &     2.32  ($\pm$0.29) & s,c \\

            5079.4 & AlO~~B$^{2}\Sigma^{+}$-X$^{2}\Sigma^{+}$~~(0,1)          &    31.41  ($\pm$0.90) &  \\

            5102.1 & AlO~~B$^{2}\Sigma^{+}$-X$^{2}\Sigma^{+}$~~(1,2)          &    17.71  ($\pm$0.92) & b \\

            5123.3 & AlO~~B$^{2}\Sigma^{+}$-X$^{2}\Sigma^{+}$~~(2,3)          &     7.83  ($\pm$0.94) &  \\

            5142.9 & AlO~~B$^{2}\Sigma^{+}$-X$^{2}\Sigma^{+}$~~(3,4)          &     1.94  ($\pm$0.57) &  \\

            5161.0 & AlO~~B$^{2}\Sigma^{+}$-X$^{2}\Sigma^{+}$~~(4,5)          &                      &  p \\

            5166.7 & TiO~~$\alpha$~~(0,0)  &    14.46  ($\pm$0.52) &  $\rceil$B  \\
            5166.3 & FeI                   &                       &   \,$\vert$B   \\
            5168.9 & FeI                   &                       &   $\rfloor$B   \\

            5228.2 & VO~~C\,$^{4}\Sigma^{-}$-X\,$^{4}\Sigma^{-}$~~(2,0) &     1.15  ($\pm$0.32) & f,c \\

            5275.8 & VO~~C\,$^{4}\Sigma^{-}$-X\,$^{4}\Sigma^{-}$~~(3,1) &     &   p \\

            5336.5 & AlO~~B$^{2}\Sigma^{+}$-X$^{2}\Sigma^{+}$~~(0,2)          &     2.07  ($\pm$0.41) & b \\

            5357.6 & AlO~~B$^{2}\Sigma^{+}$-X$^{2}\Sigma^{+}$~~(1,3)          &     2.43  ($\pm$0.88) & c \\

            5376.8 & AlO~~B$^{2}\Sigma^{+}$-X$^{2}\Sigma^{+}$~~(2,4)          &     1.61  ($\pm$1.13) & f,c \\

            5448.2 & TiO~~$\alpha$~~(0,1)      &    10.13  ($\pm$1.42) & \\

            5469.3 & VO~~C\,$^{4}\Sigma^{-}$-X\,$^{4}\Sigma^{-}$~~(1,0) &     9.84  ($\pm$1.80) & \\

            5496.4 & TiO~~$\alpha$~~(1,2)      &     7.33  ($\pm$1.40) &  \\

            5517.3 & VO~C\,$^{4}\Sigma^{-}$-X\,$^{4}\Sigma^{-}$~~(2,1) &     1.85  ($\pm$0.92) & f,c \\

            5564.3 & CrO~~B$^{5}\Pi_{-1}$-X$^{5}\Pi_{-1}$~~(2,0) &     3.79  ($\pm$0.97) &  $\rceil$B  \\ 
            5565.9 & CrO~~B$^{5}\Pi_{0}$-X$^{5}\Pi_{0}$~~(2,0)   &                       &   \,$\vert$B   \\
            5567.7 & CrO~~B$^{5}\Pi_{1}$-X$^{5}\Pi_{1}$~~(2,0)   &                       &   \,$\vert$B   \\
            5570.2 & CrO~~B$^{5}\Pi_{2}$-X$^{5}\Pi_{2}$~~(2,0)   &                       &   \,$\vert$B   \\
            5576.4 & CrO~~B$^{5}\Pi_{3}$-X$^{5}\Pi_{3}$~~(2,0)   &                       &    $\rfloor$B \\

            5736.7 & VO~~~C\,$^{4}\Sigma^{-}$-X\,$^{4}\Sigma^{-}$~~(0,0) &    10.47  ($\pm$1.14) & c \\

            5847.6 & TiO~~$\gamma'$~~(1,0) F$_{1}$-F$_{1}$ &     7.22  ($\pm$0.65) & c \\

            5872.7 & TiO~~$\gamma'$~~(1,0) F$_{2}$-F$_{2}$ &     5.14  ($\pm$0.87) &  \\ 

            5898.9 & TiO~~$\gamma'$~~(1,0) Q$_{3}$,P$_{3}$ &                        & p,B   \\

            5821.2 & TiO~~$\gamma'$~~(2,1) F$_{2}$-F$_{2}$ &     1.50  ($\pm$0.43) & \\

            5947.7 & TiO~~$\gamma'$~~(2,1) R$_{3}$         &     0.90  ($\pm$0.35) & \\

            5954.4 & TiO~~$\gamma'$~~(2,1) Q$_{3}$,P$_{3}$ &     1.74  ($\pm$0.37) & \\

            5972.2   & YO~~A$^{2}\Pi_{3/2}$-X$^{2}\Sigma^{+}$~~(0,0) &  5.24  ($\pm$0.55) & \\

            6036.2  & ScO~~A$^{2}\Pi_{3/2}$-X$^{2}\Sigma^{+}$~~(0,0) &    51.23  ($\pm$2.02) & \\

            6051.8 & CrO~~B$^{5}\Pi_{-1}$-X$^{5}\Pi_{-1}$~~(0,0) &    21.66  ($\pm$1.63) & $\rceil$B  \\ 
            6053.3 & CrO~~B$^{5}\Pi_{0}$-X$^{5}\Pi_{0}$~~(0,0) &                       &   \,|B  \\
            6054.8 & CrO~~B$^{5}\Pi_{1}$-X$^{5}\Pi_{1}$~~(0,0) &                       &   \,|B  \\
            6058.5 & CrO~~B$^{5}\Pi_{2}$-X$^{5}\Pi_{2}$~~(0,0) &                       &   \,|B  \\
            6063.5 & CrO~~B$^{5}\Pi_{3}$-X$^{5}\Pi_{3}$~~(0,0) &                       &    \,|B  \\
            6064.3 & ScO~~A$^{2}\Pi_{1/2}$-X$^{2}\Sigma^{+}$~~(0,0)~~$^{R}R_{1G}$ &    &  $\rfloor$B,p   \\
\hline
\end{tabular}
\end{center}
\end{table*}

\setcounter{table}{2}
\begin{table*}
\caption{
  Continued
}
\begin{center}
\begin{tabular}{ l l r l}
\hline
 ${\lambda}_{\rm lab}$ &  
 identification    & 
 flux~~~~~~~~~~~~~ &
 notes             \\
\hline

            6072.6 & ScO~~A$^{2}\Pi_{3/2}$-X$^{2}\Sigma^{+}$~~(1,1) &     2.44  ($\pm$0.22) & b \\

            6079.3 & ScO~~A$^{2}\Pi_{1/2}$-X$^{2}\Sigma^{+}$~~(0,0) &    74.26  ($\pm$2.49) &  $\rceil$B  \\ 
            6086.4 & VO~~C\,$^{4}\Sigma^{-}$-X\,$^{4}\Sigma^{-}$~~(0,1) &                       & $\rfloor$B   \\

            6116.0 & ScO~~A$^{2}\Pi_{1/2}$-X$^{2}\Sigma^{+}$~~(1,1) &     3.12  ($\pm$0.60) & \\

            6132.1 & YO~~A\,$^{2}\Pi_{1/2}$-X\,$^{2}\Sigma^{+}$~~(0,0) &    11.16  ($\pm$1.01) &  $\rceil$B  \\ 
            6138.8 & VO~~C\,$^{4}\Sigma^{-}$-X\,$^{4}\Sigma^{-}$~~(1,2) &                       &  $\rfloor$B   \\

            6148.7 & TiO~~$\gamma'$~~(0,0) R$_{21}$ &     1.95  ($\pm$0.28) & b \\

            6158.5 & TiO~~$\gamma'$~~(0,0) F$_{1}$-F$_{1}$  &    63.61  ($\pm$1.86) & \\

            6183.2 & TiO~~$\gamma'$~~(0,0) Q$_{32}$ &     2.37  ($\pm$0.19) & b \\

            6186.3 & TiO~~$\gamma'$~~(0,0) F$_{2}$-F$_{2}$  &    37.64  ($\pm$1.44) & \\

            6210.8 & TiO~~$\gamma'$~~(1,1) R$_{1}$  &     3.27  ($\pm$0.22) & B \\

            6214.9 & TiO~~$\gamma'$~~(0,0) F$_{3}$-F$_{3}$  &    45.47  ($\pm$1.64) & b \\

            6239.0 & TiO~~$\gamma'$~~(1,1) R$_{2}$  &     0.98  ($\pm$0.08) &  \\

            6242.2 & TiO~~$\gamma'$~~(1,1) Q$_{2}$, P$_{2}$ &     5.75  ($\pm$0.55) & \\

            6268.9 & TiO~~$\gamma'$~~(1,1) F$_{3}$-F$_{3}$  &     7.30  ($\pm$1.07) & b \\

            6321.2 & TiO~~$\gamma$~~(2,0) R$_{2}$    &     0.28  ($\pm$0.07) & f \\

            6351.3 & TiO~~$\gamma'$~~(2,2) Q$_{3}$   &     1.33  ($\pm$0.46) & c,f \\

            6394.2 & CrO~~B$^{5}\Pi_{-1}$-X$\Pi_{-1}$~~(0,1) &    33.37  ($\pm$2.00) &  $\rceil$B  \\ 
            6396.2 & CrO~~B$^{5}\Pi_{0}$-X$\Pi_{0}$~~(0,1) &                       &    \,|B   \\
            6397.8 & CrO~~B$^{5}\Pi_{1}$-X$\Pi_{1}$~~(0,1) &                       &    \,|B   \\
            6401.4 & CrO~~B$^{5}\Pi_{2}$-X$\Pi_{2}$~~(0,1) &                       &    \,|B   \\
            6407.7 & CrO~~B$^{5}\Pi_{3}$-X$\Pi_{3}$~~(0,1) &                       &    $\rfloor$B   \\

            6451.7 & CrO~~B$^{5}\Pi_{-1}$-X$\Pi_{-1}$~~(1,2) &    14.65  ($\pm$1.92) &  $\rceil$B  \\ 
            6451.9 & CrO~~B$^{5}\Pi_{0}$-X$^{5}\Pi_{0}$~~(1,2) &                       &  \,|B   \\
            6455.2 & CrO~~B$^{5}\Pi_{1}$-X$^{5}\Pi_{1}$~~(1,2) &                       &   \,|B   \\
            6459.5 & CrO~~B$^{5}\Pi_{2}$-X$^{5}\Pi_{2}$~~(1,2) &                       &   \,|B   \\
            6465.4 & CrO~~B$^{5}\Pi_{3}$-X$^{5}\Pi_{3}$~~(1,2) &                       &  $\rfloor$B   \\
            6477.8 & VO~~C\,$^{4}\Sigma^{-}$-X\,$^{4}\Sigma^{-}$~~(0,2) &     5.46  ($\pm$1.30) & b \\

            6533.3 & VO~~C\,$^{4}\Sigma^{-}$-X\,$^{4}\Sigma^{-}$~~(1,3) &    11.48  ($\pm$2.08) & b \\

            6562.6 & TiO~~$\gamma'$~~(0,1) F$_{1}$-F$_{1}$  &    24.37  ($\pm$1.02) & B \\

            6594.0 & TiO~~$\gamma'$~~(0,1) R$_{2}$ &    11.84  ($\pm$1.22) &  $\rceil$B \\ 

            6597.9 & TiO~~$\gamma'$~~(0,1) Q$_{2}$,P$_{2}$  &                       & $\rfloor$B \\

            6618.0 & TiO~~$\gamma'$~~(1,2) F$_{1}$-F$_{1}$  &     8.18  ($\pm$1.00) & c \\

            6635.4  & TiO~~$\gamma'$~~(0,1) Q$_{3}$, P$_{3}$  &     5.70  ($\pm$0.51) & \\

            6651.3 & TiO~~$\gamma$~~(1,0) F$_{3}$-F$_{3}$   &    12.07  ($\pm$0.66) & c,$\rceil$B  \\ 
            6649.8 & TiO~~$\gamma'$~~(1,2) F$_{2}$-F$_{2}$  &  & $\rfloor$B \\

            6680.8 & TiO~~$\gamma$~~(1,0) F$_{2}$-F$_{2}$   &    12.58  ($\pm$1.50) &  $\rceil$B \\ 
            6681.8 & TiO~~$\gamma'$~~(1,2) F$_{3}$-F$_{3}$  &  &  $\rfloor$B \\

            6714.5 & TiO~~$\gamma$~~(1,0) F$_{1}$-F$_{1}$   &    25.94  ($\pm$1.90) & $\rceil$B  \\ 
            6717.6 & TiO~~$\gamma$~~(2,1) F$_{3}$-F$_{3}$    &                       & $\rfloor$B   \\

            6747.6 & TiO~~$\gamma$~~(2,1) F$_{2}$-F$_{2}$   &     9.64  ($\pm$1.54) & \\

            6772.3 & CrO~~B$^{5}\Pi_{-1}$-X$^{5}\Pi_{-1}$~~(0,2) &    24.66  ($\pm$1.89) & $\rceil$B \\ 
            6774.2 & CrO~~B$^{5}\Pi_{0}$-X$^{5}\Pi_{0}$~~(0,2) &                       &  \,|B   \\
            6775.9 & CrO~~B$^{5}\Pi_{1}$-X$^{5}\Pi_{1}$~~(0,2) &                       &  \,|B   \\
            6779.6 & CrO~~B$^{5}\Pi_{2}$-X$^{5}\Pi_{2}$~~(0,2) &                       &   \,|B   \\
            6781.8 & TiO~~$\gamma$~~(2,1) F$_{1}$-F$_{1}$      &                       &   \,|B   \\
            6785.7 & CrO~~B$^{5}\Pi_{3}$-X$^{5}\Pi_{3}$~~(0-2) &                       &  $\rfloor$B   \\

            6815.1 & TiO~~$\gamma$~~(3,2)~~F$_{2}$-F$_{2}$      &     0.51  ($\pm$0.18) & f \\

            6830.7 & TiO~~b$^{1}\Pi$-X$^{3}\Pi_{1}$~~(0,0)~~R$_{12}$  &    18.24  ($\pm$2.45) & $\rceil$B  \\ 
            6836.5 & CrO~~B$^{5}\Pi_{-1}$-X$^{5}\Pi_{-1}$~~(1,3) &                     &    \,|B   \\
            6836.5 & TiO~~b$^{1}\Pi$-X$^{3}\Pi_{1}$~~(0,0)~~Q$_{12}$  &                &    \,|B   \\
            6836.6 & CrO~~B$^{5}\Pi_{0}$-X$^{5}\Pi_{0}$~~(1,3) &                       &    \,|B   \\
            6839.9 & CrO~~B$^{5}\Pi_{1}$-X$^{5}\Pi_{1}$~~(1,3) &                       &    \,|B   \\
            6844.5 & CrO~~B$^{5}\Pi_{2}$-X$^{5}\Pi_{2}$~~(1,3) &                       &    \,|B   \\
            6850.9 & CrO~~B$^{5}\Pi_{3}$-X$^{5}\Pi_{3}$~~(1,3) &                       &    $\rfloor$B   \\

            6919.0 & VO~~C\,$^{4}\Sigma^{-}$-X\,$^{4}\Sigma^{-}$~~(0,3) &     1.51  ($\pm$0.43) & f,E \\

            6976.2 & VO~~C\,$^{4}\Sigma^{-}$-X\,$^{4}\Sigma^{-}$~~(1,4) &     1.05  ($\pm$0.42) & f,c \\

\hline
\end{tabular}
\end{center}
\end{table*}

\setcounter{table}{2}
\begin{table*}
\caption{
  Continued
}
\begin{center}
\begin{tabular}{l l r l}
\hline
 ${\lambda}_{\rm lab}$ &  
 identification    & 
 flux~~~~~~~~~~~~~ &
 notes             \\
\hline

             7054.2 & TiO~~$\gamma$~~(0,0) F$_{3}$-F$_{3}$ &   119.27  ($\pm$3.68) & \\

             7087.6 & TiO~~$\gamma$~~(0,0) F$_{2}$-F$_{2}$ &   118.40  ($\pm$3.85) & \\

             7125.5 & TiO~~$\gamma$~~(0,0) F$_{1}$-F$_{1}$ &   184.10  ($\pm$3.67) &  $\rceil$B \\ 
             7124.9 & TiO~~$\gamma$~~(1,1) F$_{3}$-F$_{3}$ &  &  $\rfloor$B   \\

             7158.8 & TiO~~$\gamma$~~(1,1) F$_{2}$-F$_{2}$ &    26.39  ($\pm$3.40) & E \\

             7197.4 & TiO~~$\gamma$~~(1,1) F$_{1}$-F$_{1}$ &    33.24  ($\pm$2.42) & $\rceil$B E,s \\ 
             7196.4 & TiO~~$\gamma$~~(2,2) F$_{3}$-F$_{3}$ &   &  $\rfloor$B \\

             7270.1 & TiO~~$\gamma$~~(2,2) F$_{1}$-F$_{1}$      &     0.73  ($\pm$0.26) & c,f \\

             7343~~ & VO~~B$^{4}\Pi_{5/2}$-X$^{4}\Sigma^{-}$~~(1,0) &    14.38  ($\pm$2.06) & c \\

             7378~~ & VO~~B$^{4}\Pi_{3/2}$-X$^{4}\Sigma^{-}$~~(1,0) &    17.74  ($\pm$3.61) & c \\

             7408~~ & VO~~B$^{4}\Pi_{1/2}$-X$^{4}\Sigma^{-}$~~(1,0) &    17.23  ($\pm$4.42) & c \\

             7454~~ & VO~~B$^{4}\Pi_{-1/2}$-X$^{4}\Sigma^{-}$~~(1,0) &    21.77  ($\pm$3.94) & c \\

             7589.3 & TiO~~$\gamma$~~F$_3$-F$_3$ (0,1)  &  $>$14.01  ($\pm$0.75) & \\

             7627.7 & TiO~~$\gamma$~~F$_2$-F$_2$ (0,1)      &                      &  p,u,s  \\

             7665.8 & TiO~~$\gamma$~~F$_3$-F$_3$ (1,2)      &                      & B,u   \\

             7671.6 & TiO~~$\gamma$~~F$_1$-F$_1$ (0,1)      &                       & p,B,u   \\

             7704.9 & TiO~~$\gamma$~~F$_2$-F$_2$ (1,2)      &                       & p,B,u   \\

             7743.0 & TiO~~$\gamma$~~F$_3$-F$_3$ (2,3)  &                       &    p,B  \\
             7749.5 & TiO~~$\gamma$~~F$_1$-F$_1$ (1,2)  &    27.48   ($\pm$3.71  & \\

             7782.8 & TiO~~$\gamma$~~F$_2$-F$_2$ (2,3)  &     4.77  ($\pm$1.64) & b \\

             7828.1 & TiO~~$\gamma$~~F$_1$-F$_1$ (2,3)   &     5.69  ($\pm$1.85) & f \\

             7867.1 & VO~~B$^{4}\Pi_{5/2}$-X$^{4}\Sigma^{-}$~~(0,0)  &    87.16  ($\pm$7.38) & b \\

             7910.4 & VO~~B$^{4}\Pi_{3/2}$-X$^{4}\Sigma^{-}$~~(0,0) &    87.96  ($\pm$7.15) & b \\

             7941.9 & VO~~B$^{4}\Pi_{1/2}$-X$^{4}\Sigma^{-}$~~(0,0) &   100.888:  ($\pm$6.321) & b \\

             7982.6 & VO~~B$^{4}\Pi_{-1/2}$-X$^{4}\Sigma^{-}$~~(0,0) &   158.93  ($\pm$13.30) & b \\
 
             8030.3 & VO~~B$^{4}\Pi_{-1/2}$-X$^{4}\Sigma^{-}$~~(1,1) &    19.69  ($\pm$4.72) & \\

             8433.2 & TiO~~$\epsilon$~~(0,0)~~R$_1$             &    76.36  ($\pm$6.87) &  $\rceil$B \\ 
             8442.3 & TiO~~$\epsilon$~~(0,0)~~R$_2$,Q$_1$,P$_1$ &                       &   \,|B   \\
             8451.8 & TiO~~$\epsilon$~~(0,0)~~R$_3$,Q$_2$,P$_2$ &                       &   \,|B   \\
             8462.7 & TiO~~$\epsilon$~~(0,0)~~Q$_3$,P$_3$       &                       &  $\rfloor$B   \\

             8682.6 & VO~~B$^{4}\Pi_{-1/2}$-X$^{4}\Sigma^{-}$~~(0,1) &    33.45  ($\pm$3.67) & c \\

             8736.7 & VO~~B$^{4}\Pi_{-1/2}$-X$^{4}\Sigma^{-}$~~(1,2)~~R$_{1}$ &     2.62  ($\pm$0.66) & f \\

             8760.0 & VO~~B$^{4}\Pi_{-1/2}$-X$^{4}\Sigma^{-}$~~(1,2)~~P$_{1}$ &     7.58  ($\pm$1.70) & c \\

\hline
\end{tabular}
\end{center}
\end{table*}

\setcounter{table}{3}
\begin{table*}
\caption{
  Meaning of symbols used in Tables 2 and 3.
}
\label{notes_tab}
\begin{center}
\begin{tabular}{     r c l}
\hline
]B & - & measurements refer to the whole blend \\
c & - & uncertain continuum level \\
f & - & faint feature \\
s & - & shape-fitting was problematical \\
b & - & blended features \\
B & - & very strong blending  \\
E & - & corrected for interstellar and/or telluric absorption\\
: & - & uncertain \\
$>$& - & lower limit \\
p & - & present or probably present, but unmeasurable \\
u & - & unmeasurable (cannot be deconvolved) \\
\hline
\end{tabular}
\end{center}
\end{table*}

\end{document}